\documentclass[a4paper]{article}

\usepackage{INTERSPEECH2018}
 \usepackage{epsfig,amssymb,amsmath,multirow,boldline,graphicx,makecell,hyperref,textcomp,subfig,float,kotex}

\usepackage{xcolor}

\title{VoiceID Loss: Speech Enhancement for Speaker Verification}
\name{Suwon Shon, Hao Tang, James Glass}
\address{MIT Computer Science and Artificial Intelligence Laboratory, Cambridge, MA, USA
}
\email{\{swshon,haotang,glass\}@mit.edu}

\begin{document}

\maketitle
\begin{abstract}
In this paper, we propose VoiceID loss, a novel loss function for training a speech enhancement model to improve the robustness of speaker verification. In contrast to the commonly used loss functions for speech enhancement such as the L2 loss, the VoiceID loss is based on the feedback from a speaker verification model to generate a ratio mask. The generated ratio mask is multiplied pointwise with the original spectrogram to filter out unnecessary components for speaker verification. In the experiments, we observed that the enhancement network, after training with the VoiceID loss, is able to ignore a substantial amount of time-frequency bins, such as those dominated by noise, for verification. The resulting model consistently improves the speaker verification system on both clean and noisy conditions. 
% We further show how the enhancement is able to improve performance even in clean condition based on the frame-level cosine similarity matrix.

\end{abstract}

\section{Introduction}

By exploiting large quantities of data, especially via data augmentation~\cite{DavidSnyder2017inter,Heigold2016}, speaker embedding methods are now able to surpass the conventional i-vector~\cite{Dehak2011} approach. Many variants, largely based on multiclass classification, have been proposed to extract robust embeddings from speech. The freely available speaker recognition dataset, Voxceleb~\cite{Nagraniy2017,Chung2018}, also accelerated speaker verification improvements by having a common benchmark for different approaches to compare to.

While noise robustness is a general and hard problem for many speech processing tasks, there are relatively few studies about the use of speech enhancement for speaker recognition tasks. This is because, rather than having multiple preprocessing steps to remove noise, training speaker recognition systems using a large and diverse dataset is a simple and powerful solution.
Speaker recognition systems naturally become robust to noisy environments
when trained on a large dataset augmented with real or synthetic noise types.
This approach is especially appealing for large, overparameterized neural networks.
Besides, the objective of speech enhancement is to improve the speech quality by suppressing noise,
and makes no guarantees to downstream tasks such as speaker verification.
Even worse, the artifacts and distortions caused by speech enhancement might
even deteriorate speaker verification performance~\cite{sadjadi2010assessment}.

\begin{figure}[t]
    \centering
    \subfloat[]{\includegraphics[width=0.35\linewidth]{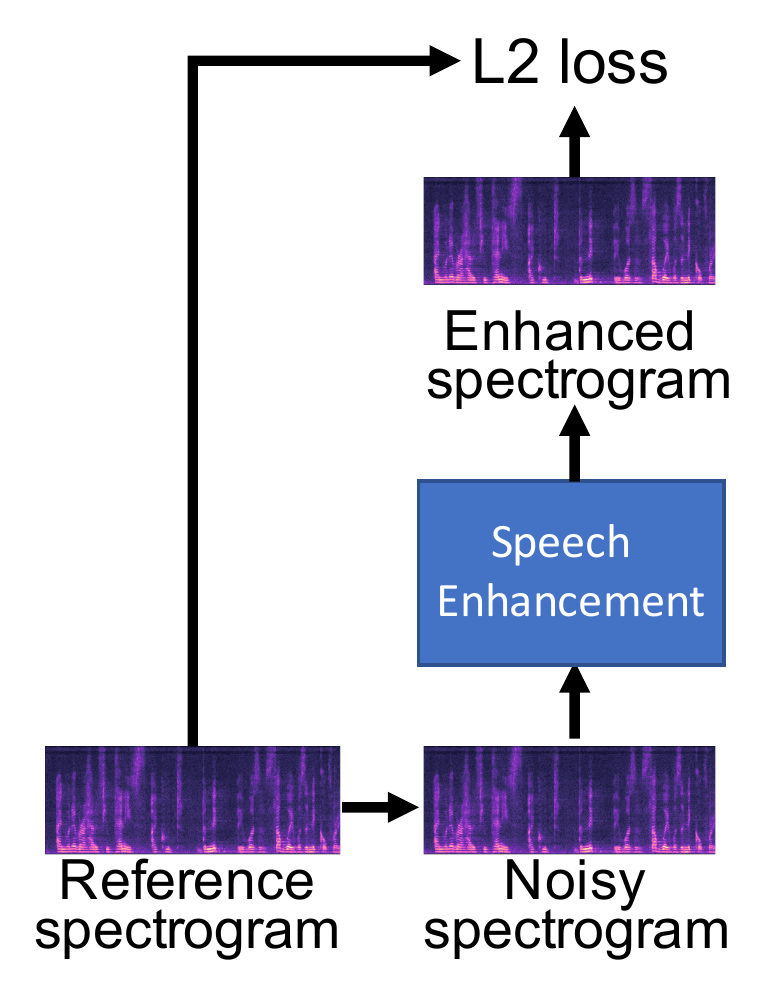}}
    \hspace{0.5cm}
    \subfloat[]{\includegraphics[width=0.55\linewidth]{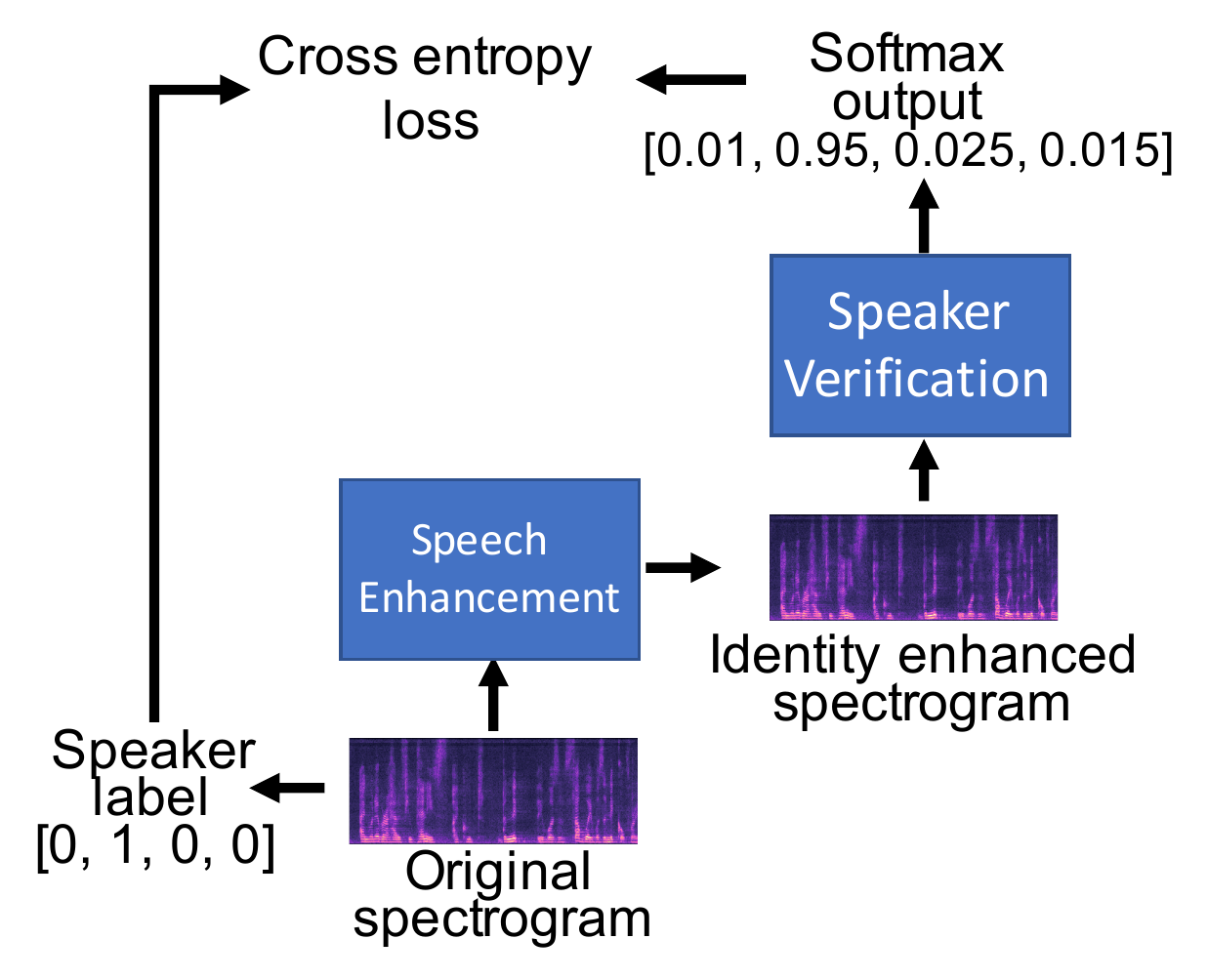}}
    \vspace{-0.3cm}
    \caption{A diagram for speech enhancement (a) using L2 loss, (b) using VoiceID loss.}
    \vspace{-0.6cm}
    \label{fig:concept}
\end{figure}

For these reasons, only a few studies have explored speech enhancement for speaker verification~\cite{ortega1996overview,sadjadi2010assessment} and most recent studies are based on the i-vector approach~\cite{Plchot2016autoencoder,Novotny2018autoencoder,Novotny2018enhance}. They used a Denoising Autoencoder (DAE) to generate an enhanced signal from the noisy signal. As shown in Figure~\ref{fig:concept}.(a), the objective of DAE is to minimize the L2 loss between the output of the model and the clean speech. During training, not only noisy-clean pairs but clean-clean pairs are also needed to prevent the DAE from deteriorating the quality of the clean signal.
These studies show improvement on the noisy and mismatched conditions, but have marginal gains on the clean and matched conditions. This is expected because the objective of DAE is to generate outputs that are closer to the inputs. 
% when it is clean speech.
Another study~\cite{Michelsanti2017} avoids deterioration from artifacts and distortions by training an individual speaker verification system using enhanced clean speech for each enhancement approach.

\begin{figure*}[ht]
    \centering
    \includegraphics[width=0.85\textwidth]{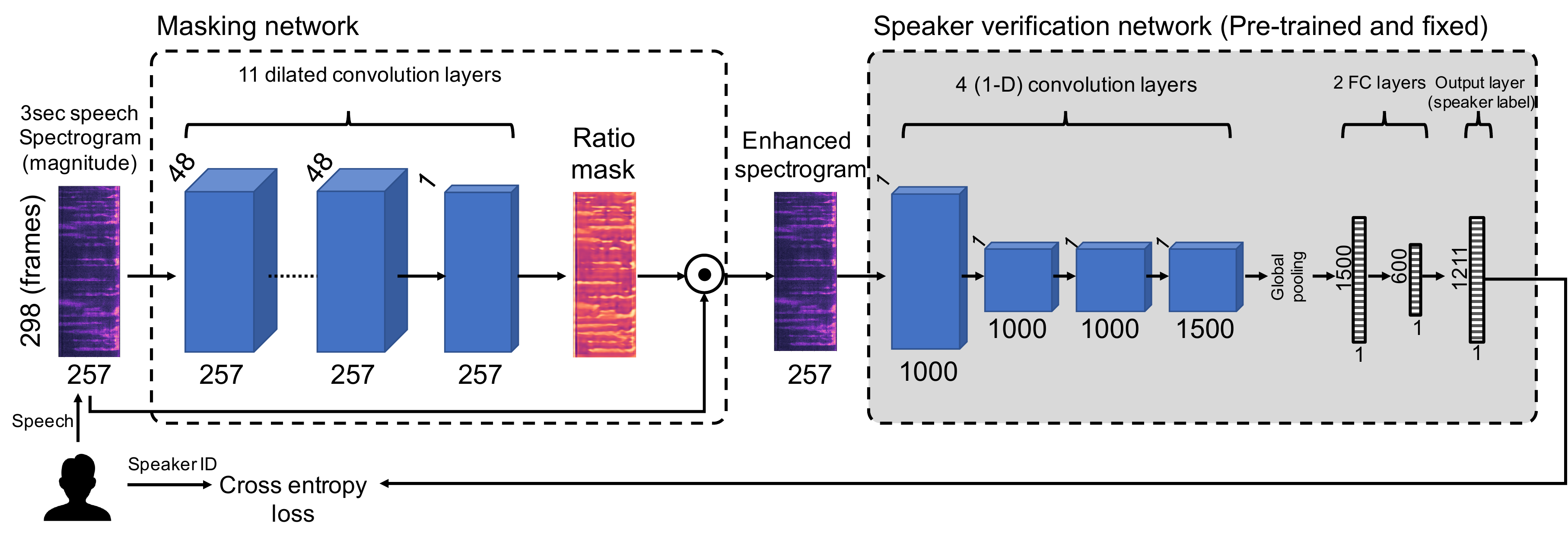}\vspace{-3mm}
    \caption{A flow chart for the VoiceID loss.}
    \vspace{-5mm}
    \label{fig:network}
\end{figure*}

To improve the effects of speech enhancement for speaker verification, we introduce a VoiceID loss that uses the error signal of a speaker verification model to train a speech enhancement model.
The overall structure of this network is shown in Figure~\ref{fig:concept}. Rather than minimizing the L2 loss between the output and the clean speech, we pass the enhanced signals to the speaker verification model, and compute the multiclass cross entropy between the output of the speaker model and the ground truth speaker label. The speech enhancement model is updated based on the cross entropy loss. Since the speaker verification system is a Deep Neural Network (DNN), the gradient can be backpropagated end to end. 

VoiceFilter~\cite{Wang2018voicefilter} is a similar approach to separate the voice of interest. They generate a ratio mask to filter out unwanted speaker's voice from the mixture of multiple speakers to obtain homogeneous speech of the target speaker. By pairing the training set with different noise types, this system could be also used for speech enhancement. However, to enable this, the system needs a strong prior, a speaker embedding, from the target speaker. Thus, the performance of separation and enhancement would heavily depend on the speaker embedding, which can be unreliable when there is only a small amount of speech from the target speaker. It is also not applicable to unseen speakers.

Another similar approach~\cite{Bagchi2018} is introduced for Automatic Speech Recognition (ASR). They have a DNN-based spectral mapper that extracts robust features from noisy speech. The spectral mapper is trained on a fidelity loss and a mimic loss. The fidelity loss is an L2 loss between the output of the spectral mapper on the clean speech and the output on the noisy speech. The mimic loss, however, is an L2 loss between the posterior probabilities of senones on the clean speech and the noisy speech. Ideally, the mimic loss should be an error between the ground truth senone class and the posterior probabilities of senones on the noisy speech. However, it is common that the ground truth alignments on the speech are not accessible, so they use the posterior probabilities on the clean speech as the ground truth. For this reason, the mimic loss might not be sufficient to minimize the word error rate on the noisy speech, and the fidelity loss is required to compensate the mismatch. This is also shown in their empirical results, where the best scale between losses is achieved with 90\% of fidelity loss and only 10\% of mimic loss.

In this paper, we only use the feedback from the verification network, i.e., the VoiceID loss. In this way, the enhancement network has the flexibility to find the important time-frequency bins for speaker verification, though the quality of speech might not be improved. From the experiment, we observed this the VoiceID loss is able to remove the noisy bins to improve the speaker verification performance on both noisy and clean condition, even for the unseen type of noise. We describe a system architecture and experimental result in subsequent sections.

\section{Speech enhancement using VoiceID loss}

The system architecture is shown in Figure~\ref{fig:network}. At first, the verification network needs to be trained using a training dataset. After training, the weights of verification network are held fixed, i.e., not updated in the subsequent steps. At the second step, the masking network and the verification network are connected to each other to form a single network. Specifically, the masking network generates a ratio mask from the input spectrogram. Then the mask is multiplied pointwise with the input spectrogram. Finally, the masked spectrogram is fed into the verification network to generate a verification output. The cross entropy loss is computed based on the output and the ground truth speaker label of the input speech.
% Then add masking network in front of the verification network. Masking network generate the ratio mask and multiplied with noisy spectrogram. This masked spectrogram is fed into the verification network. 
% For both networks, we used magnitude by applying power law compression to prevent loud audio from overwhelming soft audio.

\begin{table}[hb]
\centering
\caption{Masking network model structure.}
\label{tab:masking_configure}
\resizebox{0.7\linewidth}{!}{%
\begin{tabular}{c|c|c|c}
\hlineB{2}
Layer & Filters / Size & Dilation & Context \\ \hlineB{2}
conv1 & 48 / 1 x 7 & 1x1 & 1x7 \\ \hline
conv2 & 48 / 7 x 1 & 1x1 & 7x7 \\ \hline
conv3 & 48 / 5 x 5 & 1x1 & 11x11 \\ \hline
conv4 & 48 / 5 x 5 & 2 x 1 & 19x11 \\ \hline
conv5 & 48 / 5 x 5 & 4 x 1 & 35x11 \\ \hline
conv6 & 48 / 5 x 5 & 8 x 1 & 67x11 \\ \hline
conv7 & 48 / 5 x 5 & 1 x 1 & 71x15 \\ \hline
conv8 & 48 / 5 x 5 & 2 x 2 & 79x23 \\ \hline
conv9 & 48 / 5 x 5 & 4 x 4 & 95x39 \\ \hline
conv10 & 48 / 5 x 5 & 8 x 8 & 127x71 \\ \hline
conv11 & 1 / 1 x 1 & 1 x 1 & 127x71 \\ \hlineB{2}
\end{tabular}%
}
\end{table}

\subsection{Noisy dataset generation}
First, we generate a set of data for multiple training and test settings. We use the Voxceleb1 development set ($\mathcal{D}$) to train the networks and the test set ($\mathcal{T}$) to validate the networks. Since the dataset is collected from YouTube, the dataset is moderately noisy, but we regard the original set as the clean set. We use the noise recordings from MUSAN~\cite{Snyder2015musan} to generate corrupted versions of the Voxceleb1 development set ($\mathcal{D}^\mathcal{N}$) and the test set ($\mathcal{T}^\mathcal{N}$). Specifically, we divide the MUSAN dataset into two disjoint sets, each of which are used to augment the development and test set of Voxceleb1. We make sure we have the same types of noise in both the development and the test set, and the noise samples used to augment the test set are not seen in the development set. MUSAN consists of 4 categories of noise types: noise, music, babble, and reverberation. (The noise category contains multiple types of stationary and non-stationary noises. See~\cite{Snyder2015musan}). For the development set ($\mathcal{D}^\mathcal{N}$), we corrupt each utterance with a Signal to Noise Ratio (SNR) randomly chosen between 0 and 20 in linear scale. For the test set ($\mathcal{T}^\mathcal{N}$), we consider all four types of noise and all SNRs (in db) in the set \{0, 5, 10, 15, 20\}.

The Voxceleb1 development set ($\mathcal{D}$) has a total of 147,935 utterances from 1,211 speakers. The noise augmented dataset ($\mathcal{D}^\mathcal{N}$) is generated with same amount as the original development set ($\mathcal{D}$). The Voxceleb1 test set ($\mathcal{T}$) has 18,860 verification pairs for each positive and negative test pair, i.e., a combination of 4,715 utterances from 40 speakers. Each noisy test set ($\mathcal{T}^\mathcal{N}$) is generated with same amount as the original test set ($\mathcal{T}$).

\begin{figure}[ht]
    \centering
    \subfloat[Original]
    {\includegraphics[width=0.5\linewidth]{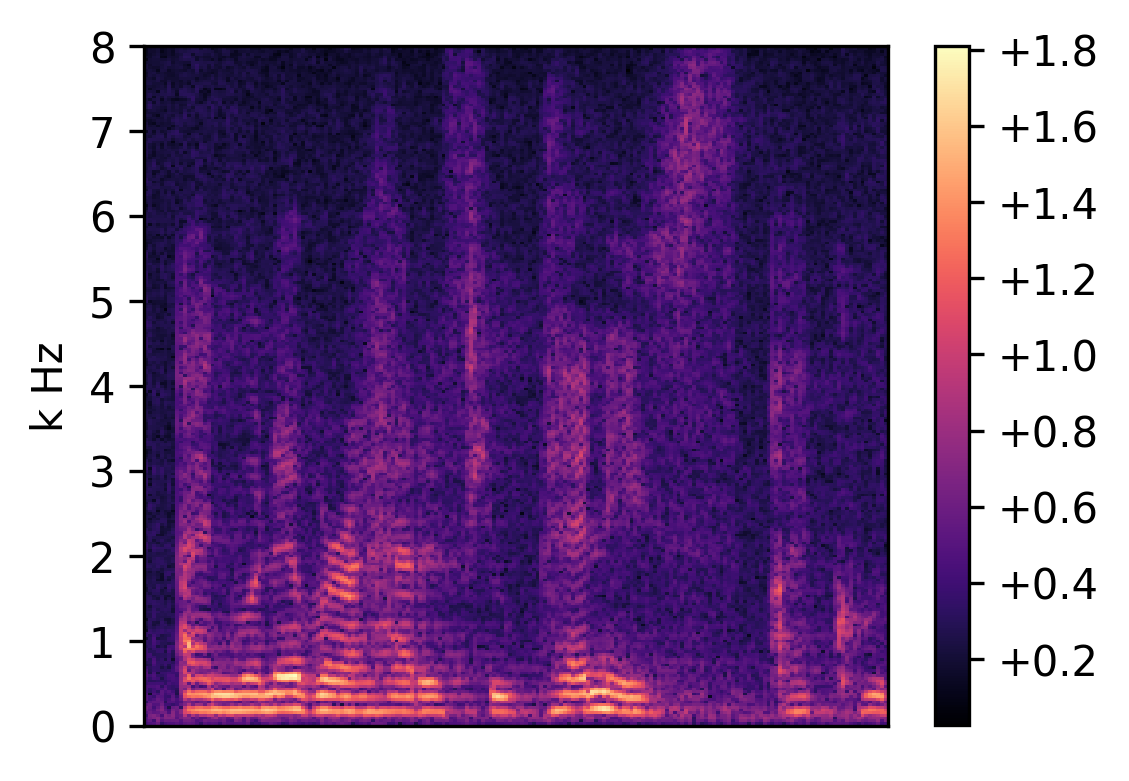}}
    % \vspace{-0.4cm}
    \subfloat[Degraded]
    {\includegraphics[width=0.5\linewidth]{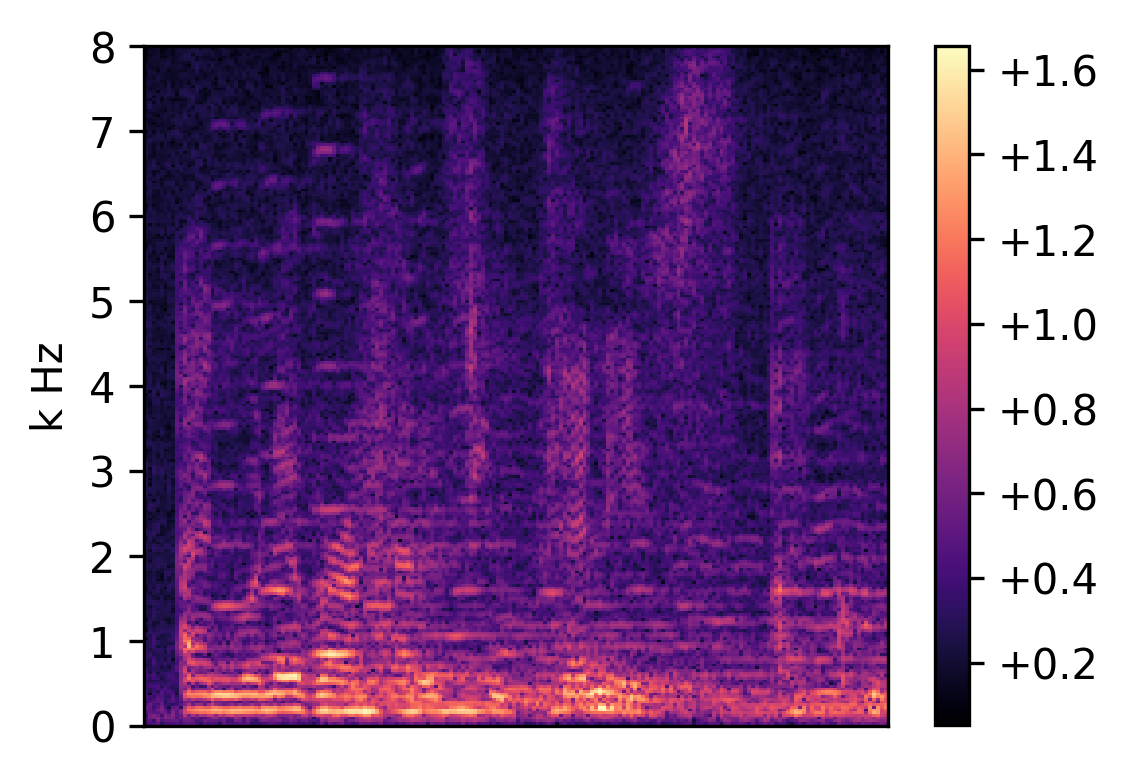}}
    \vspace{-0.4cm}
    \subfloat[Enhanced (masked)]
    {\includegraphics[width=0.5\linewidth]{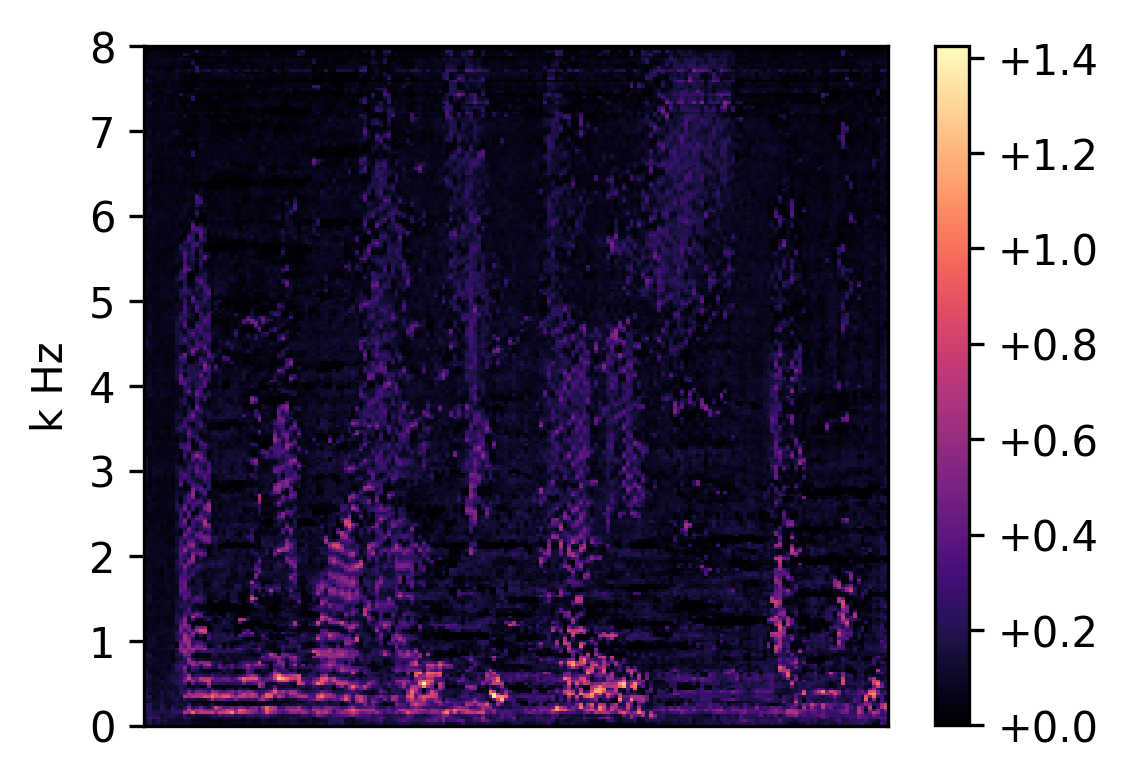}}
    % \vspace{-0.4cm}
    \subfloat[Residue]
    {\includegraphics[width=0.5\linewidth]{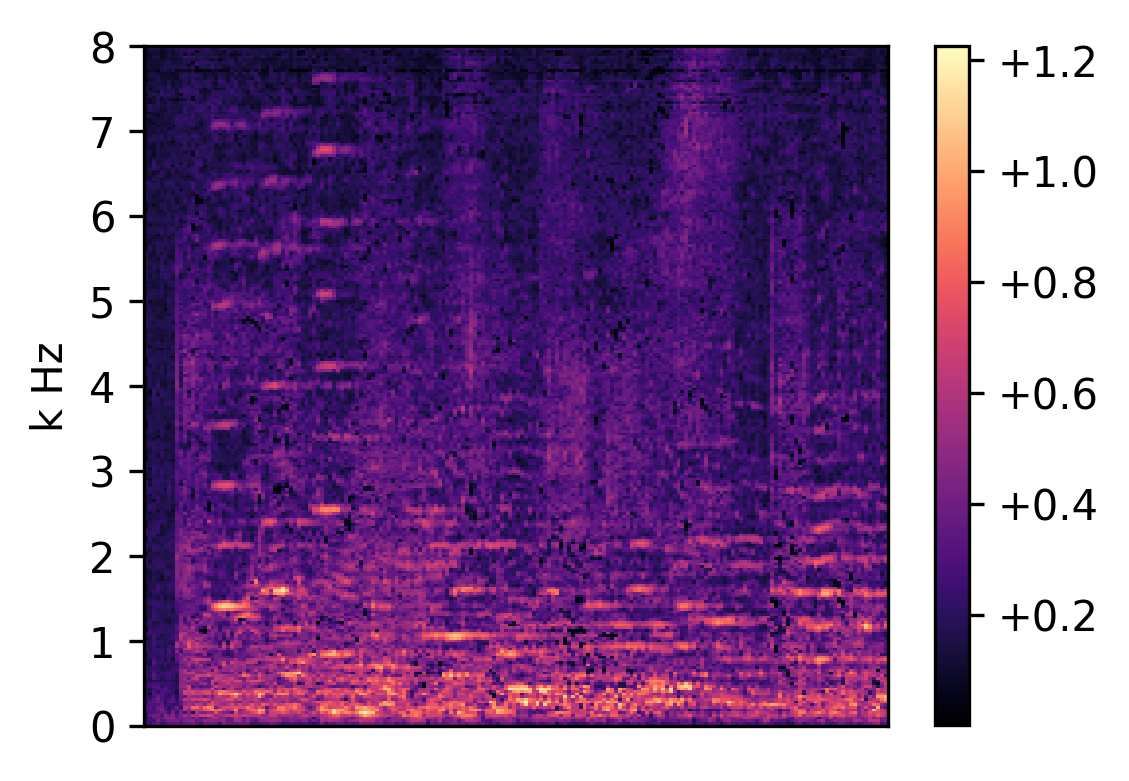}}
    \vspace{-0.4cm}
    \subfloat[Estimated mask]
    {\includegraphics[width=0.5\linewidth]{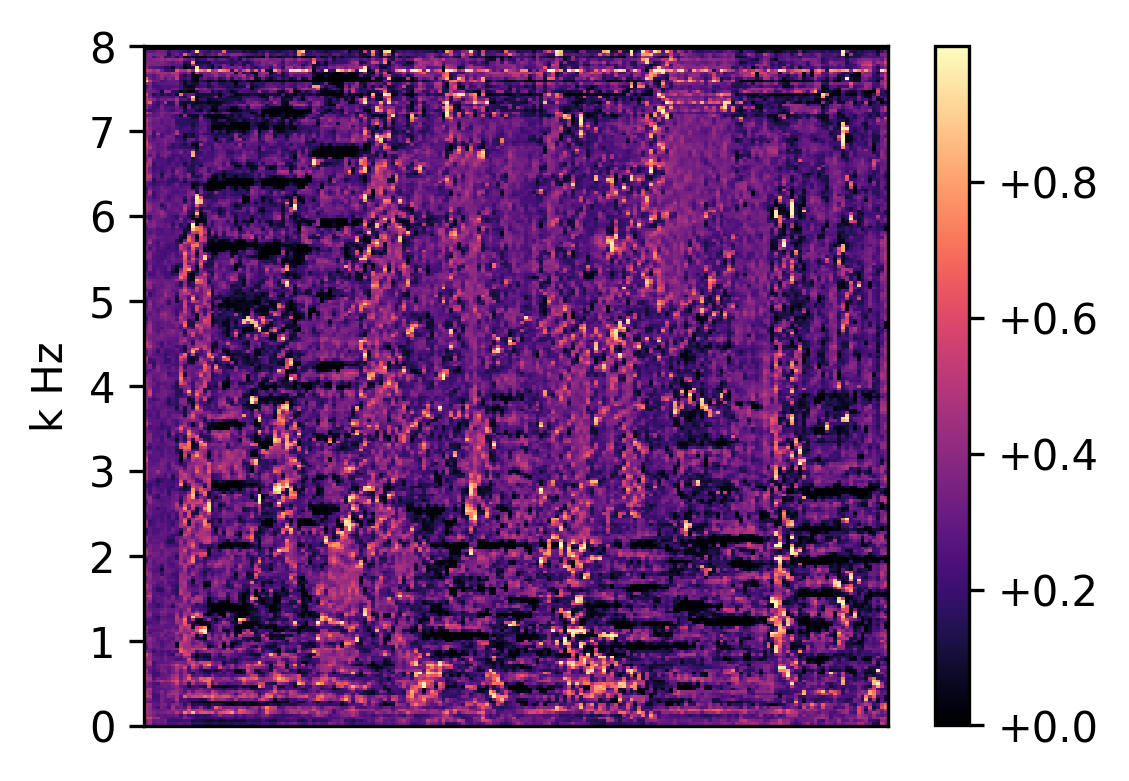}}
    \subfloat[Enhanced (DAE)]
    {\includegraphics[width=0.5\linewidth]{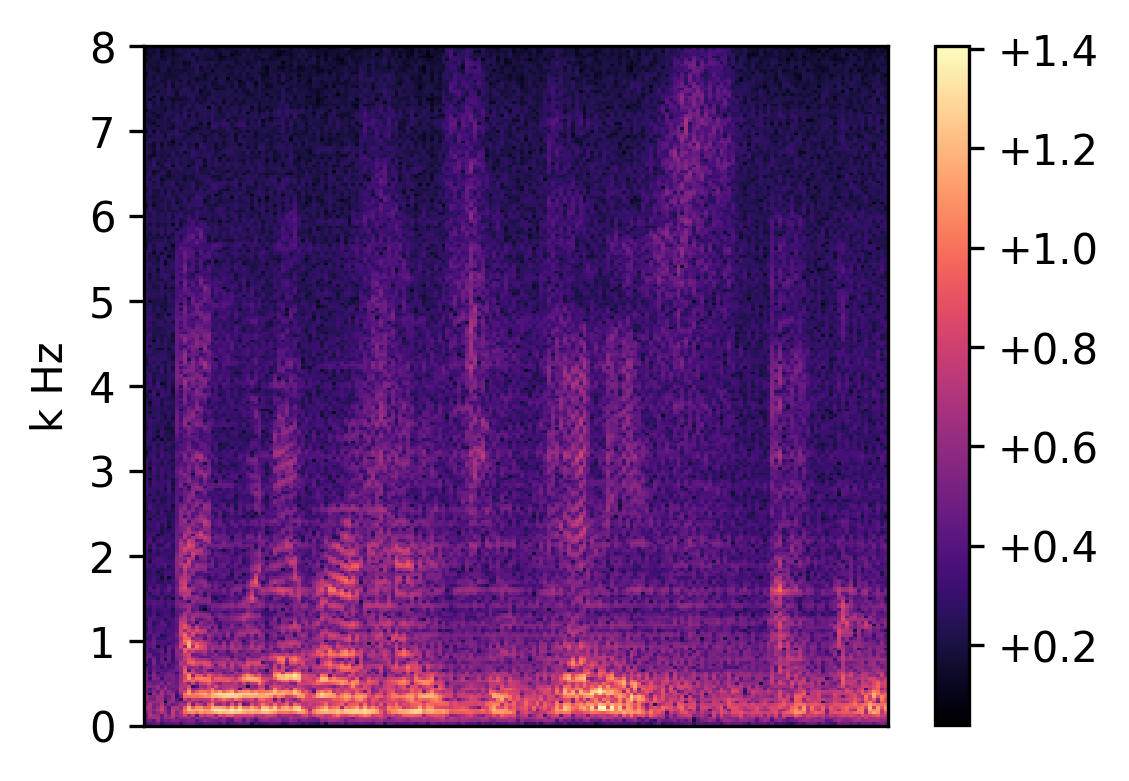}}
    % \vspace{-0.4cm}
    \caption{Example spectrograms: (a) the original spectrogram from Voxceleb1 test set, a sample never seen during training (b) a degraded sample using a musical noise with an SNR of 0 (c) the result produced by the masking network (d) the residue between the masked and the original (e) the ratio mask from the masking network (f) the result from DAE.}
    \label{fig:spec_examples}
    \vspace{-5mm}
\end{figure}

\begin{table*}
\centering
\caption{EER results obtained on the Voxceleb1 test set.}
\label{tab:exp1}
\resizebox{0.77\textwidth}{!}{%
\begin{tabular}{|c|c||cc|cc|cc||cc|cc|cc|}
\hlineB{2}
\multicolumn{2}{|c||}{Verification network}  & \multicolumn{6}{c||}{Using original set $\mathcal{D}$ } & \multicolumn{6}{c|}{Using original and augmented set $\mathcal{D}$ +$\mathcal{D}^\mathcal{N}$} \\ \hline
\multicolumn{2}{|c||}{Enhancement}  & \multicolumn{2}{c|}{-} & \multicolumn{2}{c|}{Proposed} & \multicolumn{2}{c||}{DAE} & \multicolumn{2}{c|}{-} & \multicolumn{2}{c|}{Proposed} & \multicolumn{2}{c|}{DAE} \\ \hline
\multicolumn{2}{|c||}{Enhancement network}  & \multicolumn{2}{c|}{-} &\multicolumn{2}{c|}{Using $\mathcal{D}^\mathcal{N}$} &\multicolumn{2}{c||}{Using $\mathcal{D}$+$\mathcal{D}^\mathcal{N}$} & \multicolumn{2}{c|}{-} &\multicolumn{2}{c|}{Using $\mathcal{D}$+$\mathcal{D}^\mathcal{N}$} &\multicolumn{2}{c||}{Using $\mathcal{D}$+$\mathcal{D}^\mathcal{N}$} \\ \hline
Type & SNR & EER & DCF & EER & DCF & EER & DCF & EER & DCF & EER & DCF & EER & DCF \\ \hlineB{2}
\multicolumn{2}{|c||}{Original test set $\mathcal{T}$}      & 7.73 & 0.608  & \textbf{6.99}  & \textbf{0.590} &  7.73 & 0.608 & 7.01 & 0.592 & \textbf{6.79} & \textbf{0.574} & 6.93 & 0.589 \\ \hline \hline
\multirow{5}{*}{Noise}                  & 20 & 10.34 & 0.761 & \textbf{8.10}  & \textbf{0.675} & 10.02 & 0.738 & 8.08 & 0.659 & \textbf{7.83} & \textbf{0.639} & 8.28 & 0.671 \\ 
                                        & 15 & 13.05 & 0.909 & \textbf{9.32}  & \textbf{0.699} & 11.45 & 0.833 & 8.99 & 0.720 & \textbf{8.69} & \textbf{0.686} & 8.96 & 0.761 \\ 
                                        & 10 & 17.71 & 0.987 & \textbf{11.24} & \textbf{0.770} & 14.00 & 0.943 & 10.36 & 0.770 &\textbf{9.86} & \textbf{0.747} & 10.73 & 0.869 \\ 
                                        & 5  & 24.34 & 0.999 & \textbf{14.78} & \textbf{0.885} & 18.01 & 0.988 & 12.90 & 0.851 & \textbf{12.26} & \textbf{0.830} & 13.51 & 0.958 \\ 
                                        & 0  & 31.76 & 1.000 & \textbf{20.82} & \textbf{0.983} & 23.87 & 0.998 & 17.68 & 0.945 & \textbf{16.56} & \textbf{0.938} & 18.32 & 0.994 \\ \hline \hline
\multirow{5}{*}{Music}                  & 20 & 8.97  & 0.710 & \textbf{7.54}  & \textbf{0.666} & 9.32  & 0.714 & 7.73 & 0.670 & \textbf{7.48} & \textbf{0.635} & 7.82 & 0.651\\ 
                                        & 15 & 10.60 & 0.764 & \textbf{8.23}  & \textbf{0.715} & 10.27 & 0.743 & 8.43 & 0.695 & \textbf{8.10} & \textbf{0.677} & 8.42 & 0.692 \\ 
                                        & 10 & 14.10 & 0.883 & \textbf{9.72}  & \textbf{0.760} & 11.75 & 0.808 & 9.73 & 0.760 & \textbf{9.13} & 0.733 & 9.54 & \textbf{0.728} \\ 
                                        & 5  & 20.37 & 0.992 & \textbf{13.00} & \textbf{0.819} & 15.15 & 0.941 & 12.28 & 0.833 & \textbf{11.44} & \textbf{0.818} & 11.76 & 0.846 \\ 
                                        & 0  & 29.03 & 1.000 & \textbf{18.89} & \textbf{0.937} & 20.41 & 0.993 & 17.45 & 0.935 & 16.24 & \textbf{0.913} & \textbf{15.96} & 0.961 \\ \hline \hline
\multirow{5}{*}{Babble}                 & 20 & 12.87 & 0.837 & \textbf{10.16} & \textbf{0.781} & 11.34 & 0.778 & 9.17 & 0.725 & \textbf{8.99} & \textbf{0.705} & 9.55 & 0.723 \\
                                        & 15 & 18.83 & 0.931 & \textbf{13.50} & \textbf{0.864} & 14.45 & 0.881 & 11.68 & \textbf{0.793} & \textbf{11.25} & 0.807 & 12.10 & 0.801 \\
                                        & 10 & 28.78 & 0.991 & \textbf{21.18} & \textbf{0.944} & 21.37 & 0.969 & 17.38 & \textbf{0.922} & \textbf{16.66} & 0.926 & 17.41 & 0.941\\
                                        & 5  & 38.74 & 1.000 & \textbf{33.39} & \textbf{0.996} & 33.14 & 0.997 & 28.21 & \textbf{0.992} & \textbf{27.12} & 0.996 & 29.19 & \textbf{0.992}\\ 
                                        & 0  & 44.64 & 1.000 & \textbf{42.20} & 1.000 & 43.30 & \textbf{0.999} & 38.72 & 1.000 & \textbf{37.96} & 1.000 & 41.11 & \textbf{0.999}\\ \hline \hline
\multirow{2}{*}{Reverb} & \begin{tabular}[c]{@{}c@{}}Small room\end{tabular} & 13.81 & 0.835 & \textbf{10.02} & \textbf{0.744} & 13.54 & 0.831 & 10.52 & 0.725 & \textbf{9.94} & \textbf{0.708} & 11.52 & 0.814\\ \cline{2-14} 
 & \begin{tabular}[c]{@{}c@{}}Large room\end{tabular} & 13.74  & 0.825 & \textbf{10.11} & \textbf{0.756} & 14.09 & 0.999 & 10.64 & 0.724 & \textbf{10.17} & \textbf{0.691} & 11.47 & 0.792 \\ \hline
\end{tabular}%
}
\vspace{-3mm}
\end{table*}

\subsection{Speaker Verification Network}
The speaker verification network uses 1-dimensional convolutions to consider all frequency bands at once. The network consists of 4 1D convolution layers (i.e., filters of size 40$\times$5, 1000$\times$7, 1000$\times$1, 1000$\times$1 with strides 1, 2, 1, and 1 and numbers of filters 1000, 1000, 1000, and 1500) and two fully connected (FC) layers (of size 1500 and 600). There is a global average pooling layer located between the last convolution layer and first FC layer. We use a similar structure in the previous study~\cite{Shon2018frame}, except here spectrograms are used as inputs. We use 257 frequency-bin spectrogram as input with the 25ms window size and 10ms shift to represent the speech signal. We do not use any normalization on the spectrogram. We only use the magnitudes after the short-time Fourier transform and a power-law compression with p = 0.3 (i.e., $A^{0.3}$ where $A$ is the magnitude spectrogram). In the training phase, we use a 298-frame fixed-length segment as input. We train two verification models using $\mathcal{D}$ and $\mathcal{D}$ +$\mathcal{D}^\mathcal{N}$. Multicondition training using $\mathcal{D}$ +$\mathcal{D}^\mathcal{N}$ is regarded as the standard approach to train a noise-robust speaker verification model. We extract speaker embeddings from last FC layer.

\subsection{Masking Network for Enhancement}
The masking network consists of 11 dilated convolution layers, and Table~\ref{tab:masking_configure} shows the configuration of each layer. We use the same setting for the extracting the spectrograms. To generate a ratio mask, a sigmoid function is used at the last convolution layer to have values between 0 and 1. For other layers, we use ReLUs for non-linear activations. 
Once we have the ratio mask from the last layer, the input spectrogram is multiplied with the mask and is then fed into the verification network. Training is done using the multiclass cross entropy objective with the ground truth speaker label and the verification network softmax output.
The original Voxceleb1 test set ($\mathcal{T}$) is used as a validation set. We choose the masking network that has the best Equal Error Rate (EER) on the validation set.

\section{Experiments}
We use the original Voxceleb1 test set and the augmented test set to evaluate the Voxceleb1 verification task. Cosine similarity is used to measure the score between two utterances. Performance is evaluated using EER and the Detection Cost Function (DCF). In this paper, DCF is the average of two minimum DCF scores when $P_{target}$, a priori probability of the specified target speaker, is 0.01 and 0.001.

Performance comparison is done with DAE-based speech enhancement~\cite{Plchot2016autoencoder,Novotny2018enhance,Novotny2018autoencoder}. We use an 8-layer time-delay neural network (TDNN) \cite{waibel1989phoneme} with 1000 hidden units per layer for enhancement. The architecture is the same as in \cite{tang2018study}, and the effective context size is 25 frames. We train the TDNN by minimizing the L2 loss for 10 epochs with step size 0.05 and gradient clipping of norm 5. The batch size is one utterance. After the first 10 epochs, we train the network for another 10 epochs starting from the model with the best L2 loss on the development set, with the same setting except that the step size is 0.00375 decayed by 0.75 after every epoch. The best model is chosen based on the L2 loss on the development set.

Since our proposed approach is the first to use only speaker identity for speech enhancement, we strongly recommend the readers to listen to the samples on the demo page\footnote{https://people.csail.mit.edu/swshon/supplement/voiceid-loss}. The inverse short-time Fourier transform is used to generate waveforms with the enhanced magnitudes and the original noisy phase.
Note that the objective of the proposed enhancement is for the verification network, not for a human listener. Figure~\ref{fig:spec_examples} shows example spectrograms.

\subsection{Result}

The performance (as shown in Table~\ref{tab:exp1}) is based on two types of verification model, a model trained using the clean dataset $\mathcal{D}$ and a model trained using the augmented dataset $\mathcal{D}+\mathcal{D}^\mathcal{N}$. In the clean setting, both the proposed masking approach and the DAE approach show significant improvement. The proposed masking approach shows improvement in all SNR settings consistently, but the DAE is only effective under low SNR settings. The relative improvement become marginal if we use the augmented model, as shown is the right part of the figure, but still, the proposed masking approach is effective in almost all cases. We also observe that the proposed approach shows remarkable performance under the reverberation compared to the DAE.

\begin{figure}[hb]
    \centering
    \subfloat[PESQ]
    {\includegraphics[width=0.5\linewidth]{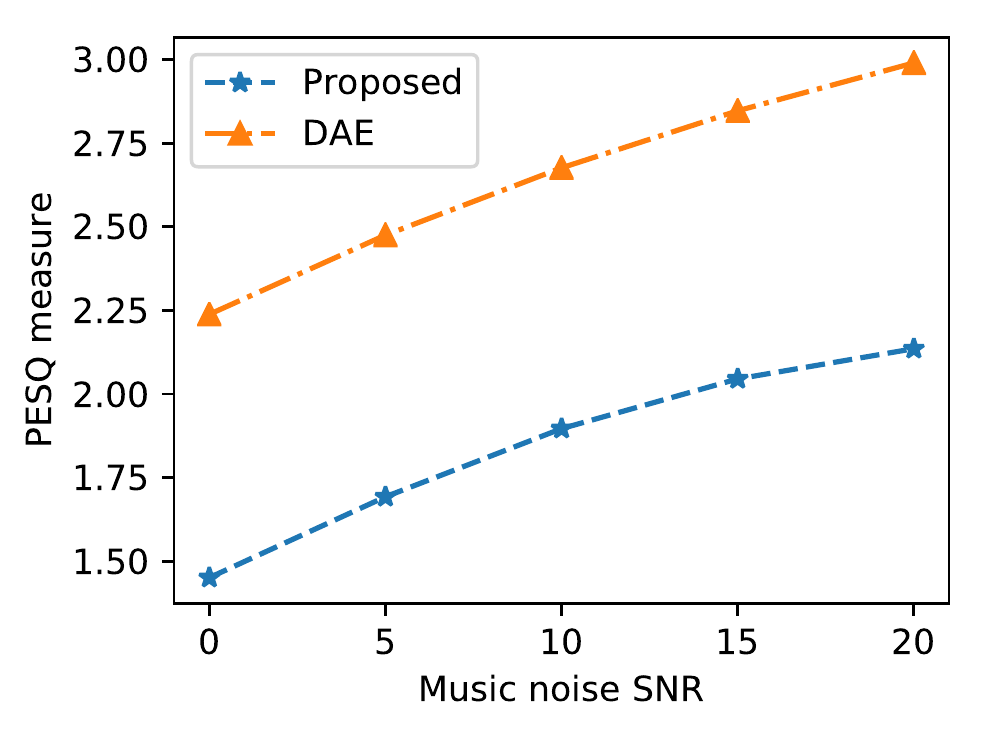}}
    % \vspace{-0.4cm}
    \subfloat[STOI]
    {\includegraphics[width=0.5\linewidth]{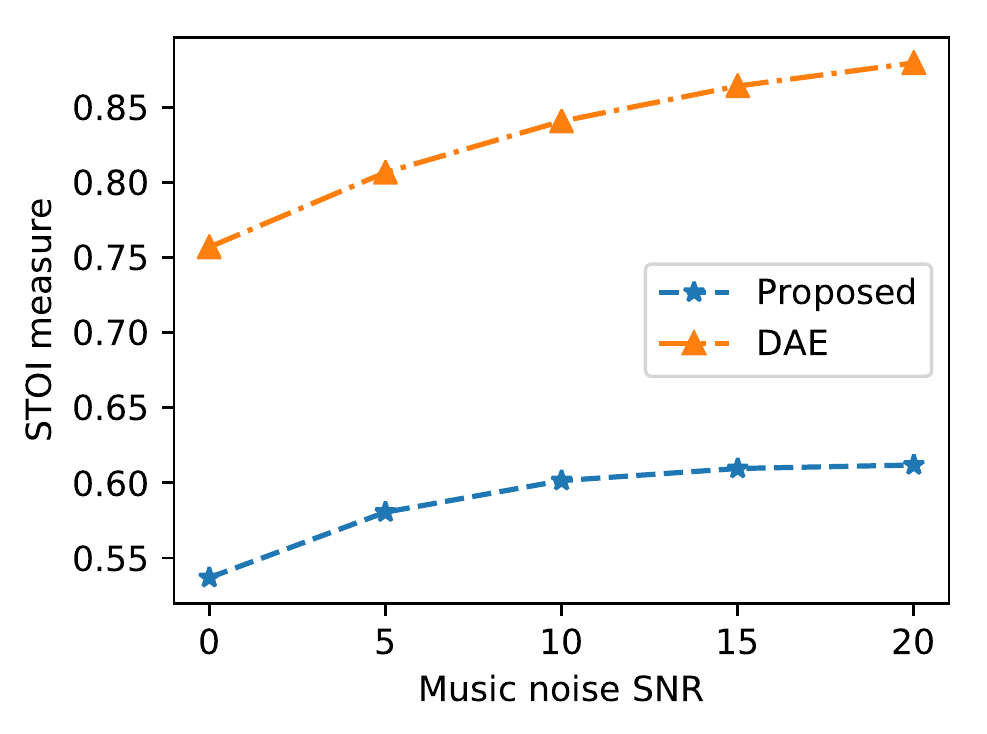}}
    % \vspace{-0.4cm}
    \caption{Speech quality measures after enhancement comparing the proposed approach and the DAE.}
    \label{fig:objective_measure}
    \vspace{-3mm}
\end{figure}

\begin{table}
\centering
\caption{Performance under unseen noise type (musical noise). Both verification and mask network trained noise augmented development set without any musical noise ($\mathcal{D^{N-M}}$).}
\resizebox{0.45\textwidth}{!}{%
\begin{tabular}{|c|c|c|c|c|c|}
\hline
\multicolumn{2}{|c|}{Verification network} & \multicolumn{2}{c|}{Using $\mathcal{D}$} & \multicolumn{2}{c|}{Using $\mathcal{D}+\mathcal{D^{N-M}}$ } \\ \hline
\multicolumn{1}{|c|}{\begin{tabular}[c]{@{}c@{}}Unseen \\ Noise\\ SNR\end{tabular}} & \multicolumn{1}{c|}{Enhancement} & EER & DCF & EER & DCF \\ \hlineB{2}

\multirow{3}{*}{20} & - & 8.97  & 0.710 & 8.00 & 0.702  \\ \cline{2-6} 
 & Proposed & \textbf{7.94} & \textbf{0.701} & \textbf{7.76} & \textbf{0.691} \\ \cline{2-6} 
 & DAE & 9.53 & 0.729 & 7.90 & 0.673 \\ \hline\hline

\multirow{3}{*}{15} & - & 10.60 & 0.764 & 8.66 & 0.732 \\ \cline{2-6} 
 & Proposed & \textbf{8.74} & \textbf{0.747} &\textbf{8.49} & \textbf{0.712} \\ \cline{2-6} 
 & DAE & 10.49 & 0.777 & 8.52 & 0.732 \\ \hline\hline

\multirow{3}{*}{10} & - & 14.10 & 0.883 & 10.32 & 0.774 \\ \cline{2-6} 
 & Proposed & \textbf{10.62} & \textbf{0.811} & 9.95 & \textbf{0.766} \\ \cline{2-6} 
 & DAE & 12.76 & 0.847 & \textbf{9.93} & 0.771 \\ \hline\hline

\multirow{3}{*}{5} & - & 20.37 & 0.992 & 13.57 & 0.872 \\ \cline{2-6} 
 & Proposed & \textbf{14.19} & \textbf{0.864} & \textbf{12.39} & 0.865 \\ \cline{2-6} 
 & DAE & 16.92 & 0.953 & 13.05 & \textbf{0.846} \\ \hline\hline

\multirow{3}{*}{0} & - & 29.03 & 1.000 & 19.73 & 0.970 \\ \cline{2-6} 
 & Proposed & \textbf{21.00} & \textbf{0.965}  & \textbf{17.28 } & 0.958  \\ \cline{2-6} 
 & DAE & 24.01 & 0.998 & 18.94 & \textbf{0.955} \\ \hline
\end{tabular}%
}
\label{tab:exp2}
\end{table}

Table~\ref{tab:exp2} shows the performance of models under unseen noise types. We exclude the musical noise from the augmented development set ($\mathcal{D^{N-M}}$). In this case, both the masking and verification networks are not exposed to musical noise during training. For the setting of unseen noise, the proposed approach also shows better performance than the DAE.

Objective measures for speech enhancement quality such as Perceptual Evaluation of Speech Quality (PESQ) and Short-Time Objective Intelligibility (STOI) are used for comparison in Figure~\ref{fig:objective_measure}. As expected, we observe that better speech quality does not imply better speaker verification. This indicates that speech enhancement should be customized to the eventual downstream task for maximum effectiveness.

\subsection{Discussion}
\begin{figure}[ht]
    \centering
    \subfloat[Original]
    {\includegraphics[width=0.465\linewidth]{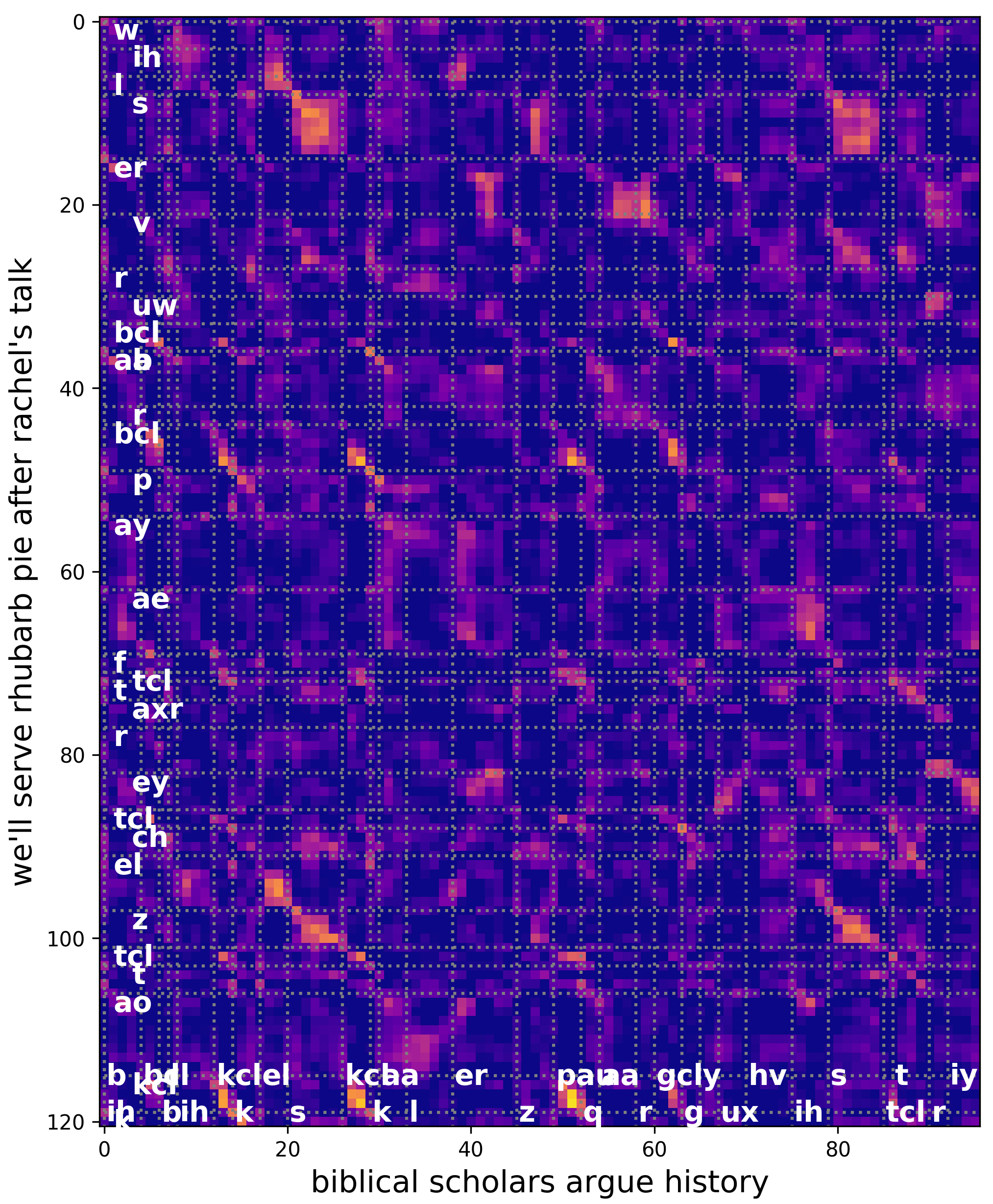}}
    % \subfloat[Noisy]
    % {\includegraphics[width=0.315\linewidth]{faem0_sx402_sx42_music0_trimmed.png}}
    \subfloat[After enhancement]
    {\includegraphics[width=0.54\linewidth]{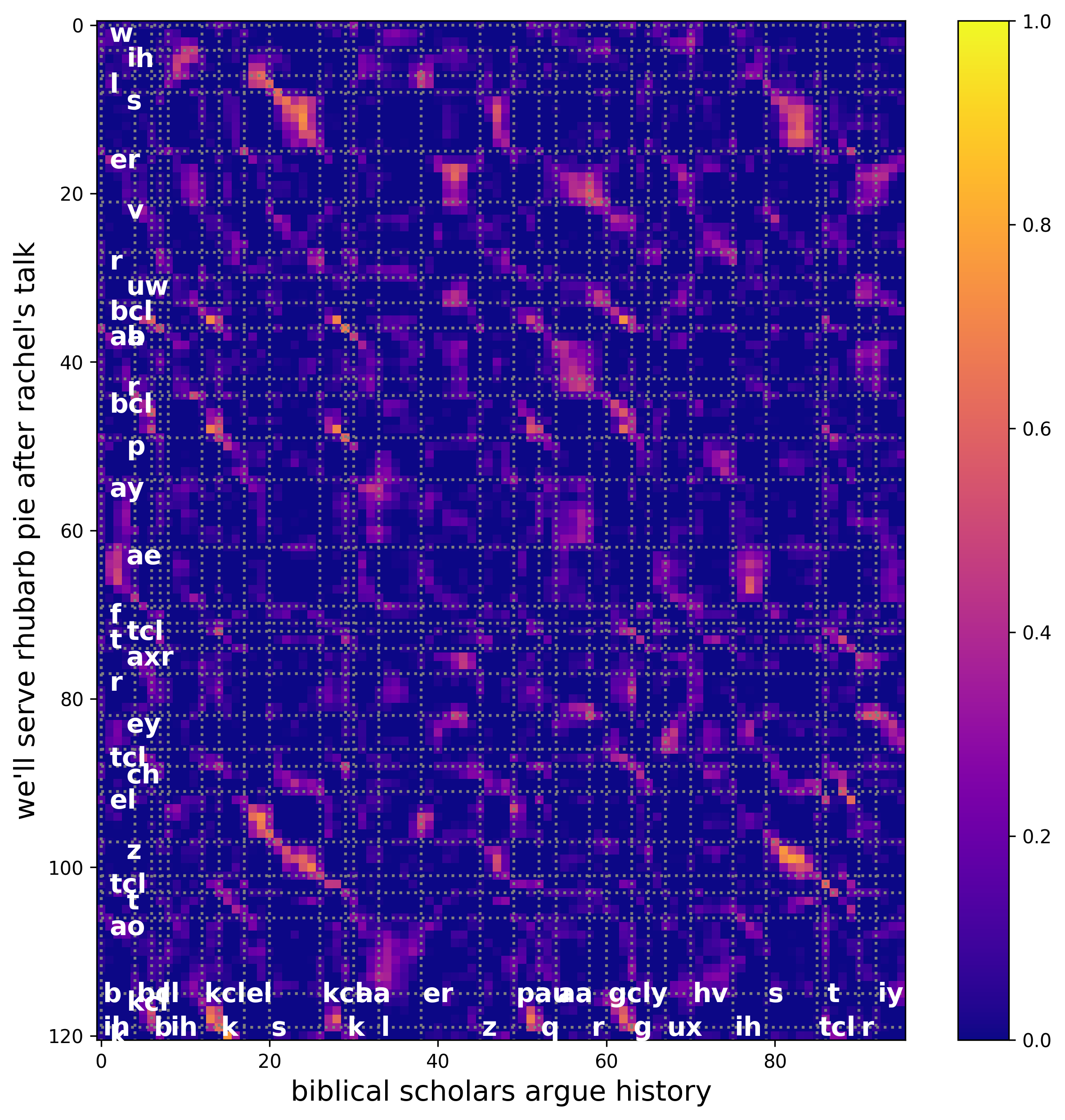}}
    \vspace{-0.2cm}
    \caption{A frame-level cosine similarity matrix between two sentences spoken by same person in TIMIT.}
    \label{fig:framelevel}
    \vspace{-5mm}
\end{figure}

Interestingly, the proposed approach improves performance even in the clean setting for both verification models. This means that the VoiceID loss removes not only the noise but also unnecessary time-frequency bins from the spectrograms. The improvement also shows in the frame-level cosine similarity matrix, as shown in Figure~\ref{fig:framelevel}. The analysis approach of using frame-level matrix is first introduced in~\cite{Shon2018frame}. We follow the same approach to compute the matrix before and after enhancement using two utterances from the same speaker in the TIMIT dataset, a clean and studio-level dataset. In Figure~\ref{fig:framelevel} (a), we see low scores between the different phonemes and high scores between the same phonemes. These low scores become close to 0 after enhancement. We hypothesize that the ambiguity of the different phonemes is removed by the mask and only similar phonemes with strong similarity have high scores.

A limitation of this study is that all experiments are done based on the speaker verification system in the previous study~\cite{Shon2018frame}. We do not use Angular Softmax~\cite{Cai2018,Hajibabaei2018}, Probabilistic Linear Discriminant Analysis (PLDA), ResNet~\cite{he2016deep,Villalba2019state}, and various utterance aggregation approaches~\cite{Xie2019,Cai2018,Okabe2018}, which show better performance than the Softmax and Cosine similarity back-end. Also, we do not consider acoustic features such as log-Mel filter-banks. We believe these variants would give more robustness in overall performance with similar margin with and without enhancement. In the future, we will consider a further study thoroughly on the use of speech enhancement with the cutting-edge verification system.

\section{Conclusion}

Motivated by the discrepancy in objectives between speech enhancement and speaker verification, we propose a novel speech enhancement approach using the VoiceID loss to improve speaker verification. The proposed approach uses speaker identity information directly to generate ratio masks for emphasizing voice characteristics and filtering out unnecessary time-frequency bins, such as noise or even speech that does not carry strong voice characteristics. Experimental results show the effectiveness of the proposed approach combined with the speaker verification model in both clean and noisy settings.

% This approach can be also utilized for speaker-specific or device-specific speech enhancement without domain adaptation or re-training of speaker verification network. In the future, we will address this by composing dataset properly. Furthermore, we will investigate how the proposed approach affects the speech recognition task. 

\clearpage
\newpage
\bibliographystyle{IEEEtran}
\bibliography{mybib}

\end{document}